\UseRawInputEncoding
\documentclass[twocolumn,superscriptaddress,secnumarabic,amssymb,nobibnotes,aps,prl]{revtex4-1}

\usepackage{graphicx}
\usepackage{epstopdf}
\usepackage{dcolumn}
\usepackage{bm}
\usepackage{amssymb,amsmath}

\begin{document}

\title{Narrow-band photodetection by heterolattice electronic
transitions in TiO$_2$ glass embedded with nanocrystals} %

\author{Zhipeng Wei}
\affiliation{State Key Laboratory of Luminescence and
Applications, Changchun Institute of Optics, Fine Mechanics and
Physics, Chinese Academy of Sciences, No.3888 Dongnanhu Road,
Changchun, 130033, People's Republic of China
}%

\affiliation{%
State Key Laboratory of High Power Semiconductor Lasers, Changchun University of Science and Technology, \\
7089 Wei-Xing Road, Changchun 130022, People¡¯s Republic of China
}%

\author{Xuan Fang}

\affiliation{%
State Key Laboratory of High Power Semiconductor Lasers, Changchun University of Science and Technology, \\
7089 Wei-Xing Road, Changchun 130022, People¡¯s Republic of China
}%

\author{Peter Y. Yu}

\affiliation{%
Department of Physics, University of California at Berkeley, Berkeley, CA 94720, USA
}%

\author{Jiaxu Yan}

\affiliation{State Key Laboratory of Luminescence and
Applications, Changchun Institute of Optics, Fine Mechanics and
Physics, Chinese Academy of Sciences, No.3888 Dongnanhu Road,
Changchun, 130033, People's Republic of China
}%

\author{Dengkui Wang}
\author{Jilong Tang}

\affiliation{%
State Key Laboratory of High Power Semiconductor Lasers, Changchun University of Science and Technology, \\
7089 Wei-Xing Road, Changchun 130022, People¡¯s Republic of China
}%

\author{Kewei Liu}
\author{Hai Xu}

\affiliation{State Key Laboratory of Luminescence and
Applications, Changchun Institute of Optics, Fine Mechanics and
Physics, Chinese Academy of Sciences, No.3888 Dongnanhu Road,
Changchun, 130033, People's Republic of China
}%

\author{Xaiobo Chen}

\affiliation{%
Department of Chemistry, University of Missouri $\textendash$ Kansas City, Kansas City, Missouri 64110, USA.
}%

\author{Lei Liu}
\email{liulei@ciomp.ac.cn} 

\affiliation{State Key Laboratory of Luminescence and
Applications, Changchun Institute of Optics, Fine Mechanics and
Physics, Chinese Academy of Sciences, No.3888 Dongnanhu Road,
Changchun, 130033, People's Republic of China
}%

\author{D. Z. Shen}
\email{shendz@ciomp.ac.cn} 
\affiliation{State Key Laboratory of Luminescence and
Applications, Changchun Institute of Optics, Fine Mechanics and
Physics, Chinese Academy of Sciences, No.3888 Dongnanhu Road,
Changchun, 130033, People's Republic of China
}%

\date{\today}%

\begin{abstract}

In this letter, we propose that the narrowband photodetection can be realized in the skin layer of semiconductors with the spatially separated photo-excited carriers from the nanoscale heterolattice electronic transitions (NanoHLETs). The NanoHLET photodetection is demonstrated by measuring the photoconductive responses of the $\sim$ 1 $\mu$m thick films of TiO$_{2}$ glass embedded with rutile and anatase TiO$_{2}$ nanocrystals. As containing only rutile nanocrystals, the TiO$_{2}$ glass film presents the responsivity curve with only one sharp peak centered at 423.0 nm with a full-width-at-half-maximum (FWHM) of 13.7 nm. This NanoHLET mechanism may open a new way in making ultra-small narrow-band photodetectors matching the current nanoscale electronic technology.

\end{abstract}

\maketitle

Photodetectors that convert optical signals into electrical ones are essential components in modern optoelectronic devices. For the purpose of matching them with those electronic technologies currently in nanoscale \cite{N1,N2,N3}, it is particularly desirable to scale down the size of photodetectors to the nanoscale as well. By now, photodetectors have been achieved well in the submicrometer scale in many ways, such as nanowire avalanche photodiodes\cite{N1, N4},  metal$\textendash$semiconductor$\textendash$metal photodetectors with antenna resonance\cite{N2}, superconducting nanowire detectors\cite{N5}, and so on\cite{N6,N7,N8,N9,N10,N11}. However, these nanodetectors normally cannot discriminate the photon frequency well, as they respond to photons over a broad frequency band starting from the absorption edge. Thus, they cannot be used directly in those detection applications requiring narrow-band response, sharp frequency discrimination, or accurate optical signal processing. In making narrow-band photodetectors, the common method is to equip the normal broadband ones with additional light frequency-selecting elements consisting of optical filters, prisms, gratings, or interferometers, etc. Obviously, these additional components may result in undesirable effects, such as increased volume, optical loss, complexity in architecture and fabrication, limited pixel density for imaging, increased cost, and interference effect etc.

For compactness, the narrow-band photodetectors can be also built without external filters, in the so called `self-filtering'\cite{N12, N13}, `filter-free'\cite{N14}, or `filterless'\cite{N15} ways. A typical self-filtering narrow-band photodetector was identified, in 1960s, as a p-n junction having a homogeneous direct-gap semiconductor base, sufficiently thick (more than 70 $\mu$m), to consume out those photons with energy beyond threshold \cite{N12, N13}. It has been revealed that for such a semiconductor those photons with energy slightly higher than bandgap will be absorbed mostly in the near-surface region\cite{N13}.  And with the active surface recombination states, the electron-hole pairs formed near the surface will not contribute to the photoconduction process\cite{N13}. In organic materials, the filterless narrowband photodetection was early recognized in the MEH-PPV films in a sandwich cell geometry between electrodes of ITO and semitransparent aluminum\cite{N16}. However, these devices were limited by the excitonic nature of the organic semiconductors, which severely affects charge generation and hence transport and external quantum efficiency\cite{N15}. Recently, Armin et al. have aroused great interest in a new concept they proposed, i.e. charge collection narrowing (CCN), to realize narrowband spectral response with broadband absorbing organic semiconductors\cite{N15,N17,N18,N19}. The CCN mechanism lies in the fact that, in the electro-optics of thick junctions (thickness $\textgreater$$\mu$m), all photons with energy greater than the optical gap are absorbed near the surface of the transparent anode, while those at the optical gap, which have a much lower extinction coefficient, can propagate through the cavity\cite{N15}. While the surface-generated carriers will be subject to significant recombination losses\cite{N18}, only those carriers generated within the volume experience a more balanced electron and hole transport\cite{N15}, resulting in the narrow-band photoresponse. Taking $\sim$1-mm-thick single crystals of hybrid perovskite semiconductors as the photoactive materials, Fang $et$ $al$. have successfully fabricated the ultra-narrowband-response photodiodes which has a very narrow spectral response with FWHM of $\textless$20 nm\cite{N15,N20}. Their narrowband photodetection has also been explained by the strong surface-charge recombination of the excess carriers close to the crystal surfaces generated by short-wavelength light\cite{N20}. And very recently, by adjusting the surface-charge recombination, Xue et al. \cite{N21}  have achieved ultra-narrow response photodetector with FWHM of $\sim$12 nm  in the inorganic CsPbX$_3$ polycrystalline perovskite films with thickness of $\sim$20 $\mu$m.

So far, the self-filtering narrowband photodetectors, have been developed for over half a century\cite{N12}. By now, the efficient mechanism of narrowband photodetection has been identified, either in inorganic or organic self-filtering detectors as discussed above, as based on sacrificing the absorbed photons in the near surface region where the strong charge recombination occurs and only those photons penetrate well below surface can induce photocurrent effectively. However, such near-surface filtering will still result in certain loss of input photons, and more importantly, that makes it hard to miniaturize photodetectors further into the submicron scale to match the nanoscale electronic devices.  Here, we propose that the narrow-band photodetection can be directly realized in the near surface region of semiconductors through the NanoHLETs.  This new approach is demonstrated by making a narrow-band photoconductive TiO$_{2}$, enabled by the photo-excited electronic transitions the valence band maximum (VBM) of glass to the conduction band minimum (CBM) of nanocrystals.

\begin{figure}
\includegraphics[width=0.45\textwidth,clip]{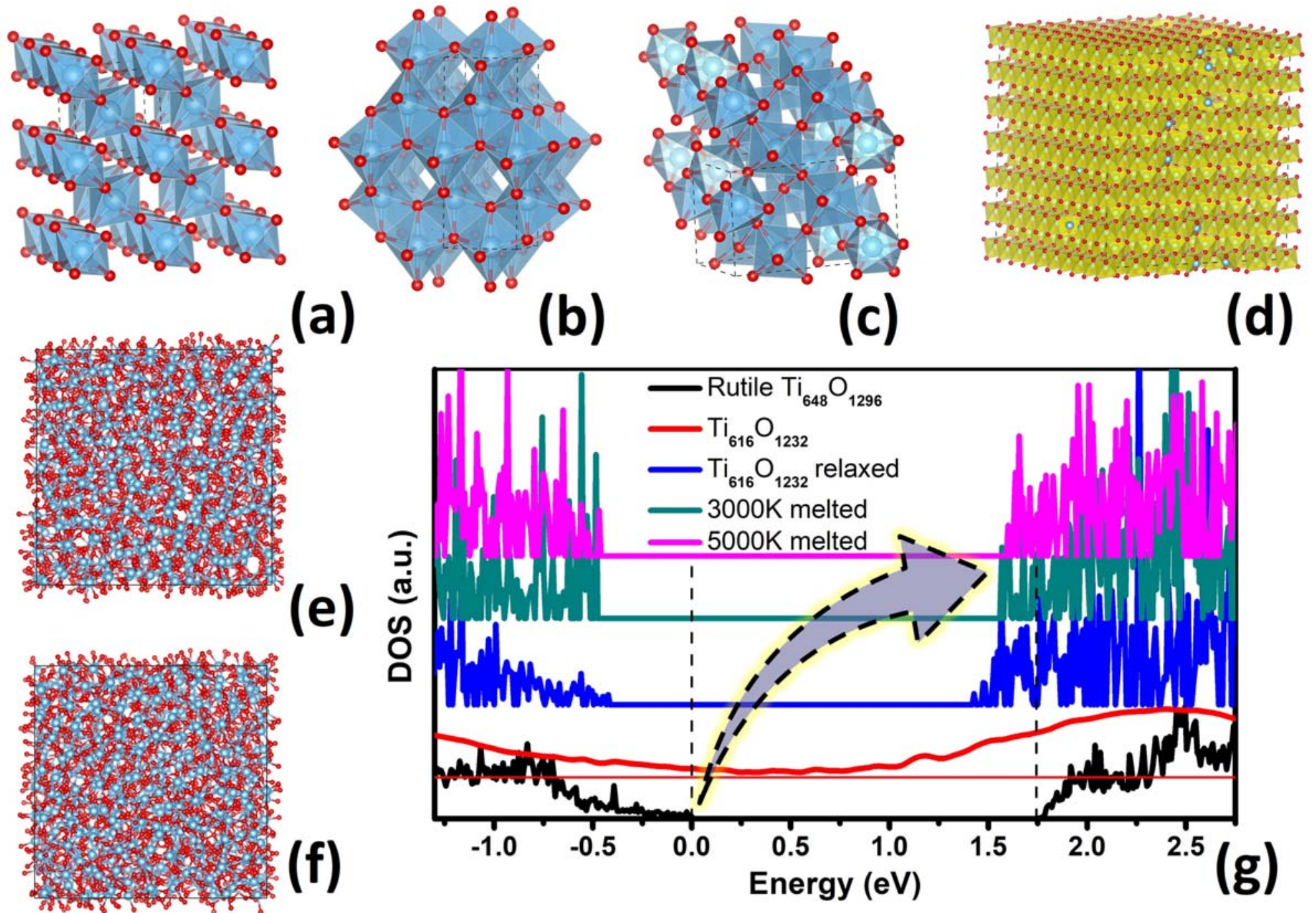}
\caption{\label{fig:figure1} The lattices of rutile (a), anatase (b), brookite (c), the Ti$_{616}$O$_{1232}$ model (d)
  where 5$\%$ TiO$_{2}$ units are removed randomly from with a (6$\times$6$\times$9) rutile supercell
  Ti$_{648}$O$_{1296}$, the 3000K-Ti$_{616}$O$_{1232}$ glass model (e), and the 5000K-Ti$_{616}$O$_{1232}$
  glass model (f); and the calculated DOS curves (g).
  Here the sky blue and red balls represent Ti and O atoms, respectively, and energy zero is set to the VBM of bulk rutile TiO$_{2}$.}
\end{figure}

The NanoHLET concept proposed here originates from the structural and electronic features of disordered TiO$_{2}$. As a common oxide semiconductor, TiO$_{2}$ has quite a few bonding styles\cite{N22}, such as tetragonal rutile (Fig.\ref{fig:figure1}(a)), tetragonal anatase (Fig.\ref{fig:figure1}(b)), orthorhombic brookite (Fig.\ref{fig:figure1}(c)), and monoclinic TiO$_{2}$(B), and that shifts their electronic bands relatively to certain extent. Here, we inspect the changes the electronic bands of TiO$_{2}$ upon lattice disorder using first-principle density functional theory calculations. Starting from a (6$\times$6$\times$9) rutile supercell with formula Ti$_{648}$O$_{1296}$, we built a disordered Ti$_{616}$O$_{1232}$ by randomly removing 5$\%$ of TiO$_{2}$ units from the bulk Ti$_{648}$O$_{1296}$ supercell.
As shown in Fig.\ref{fig:figure1}(d), this Ti$_{616}$O$_{1232}$ model is characterized by broken bonds generated by the 64 TiO$_{2}$-vacancies selected randomly. Fig. 1g compares the calculated density-of-states (DOS) of the band structures of these two models, which shows the dangling broken bonds of Ti$_{616}$O$_{1232}$ create a lot of localized states within the forbidden band gap of the bulk rutile. These gap states as in the red curve are so numerous that they merge into a quasi-continuum. But these gap states can be totally removed by enough structural relaxation as shown by the blue calculated DOS curves in Fig.\ref{fig:figure1}(g). That can be ascribed to the `healing' capability of TiO$_{2}$ in reconstructing its bonding structure around the vacancy `wounds', by contracting TiO bonds further due to the unsaturated coordination. These stronger TiO bonds move down the electronic bands overall, whose VBM and CBM stays about 0.4 eV below those of bulk rutile.

While the Ti$_{616}$O$_{1232}$ model is built arbitrarily, it would be more instructive to examine the realistic lattice disorders in glass-form TiO$_{2}$. We have built theoretically two TiO$_{2}$ glass models by quenching a Ti$_{616}$O$_{1232}$ melt from 3000K and 5000K, respectively. The results are shown in Fig.\ref{fig:figure1}(e) and Fig.\ref{fig:figure1}(f), respectively. Both models show a highly disordered lattice containing voids and no periodicity. Their corresponding DOS are presented by the dark cyan and magenta curves in Fig.\ref{fig:figure1}(g).  Similarly, lower than the bulk ones stay their electronic bands, and in particularly their CBM stays about 0.2 eV below that of bulk rutile. That indeed hints that the electronic transition from rutile VBM to the glass CBM would be different from others if excited by photons as indicated by the grey arrow in Fig.\ref{fig:figure1}(g). That is, limited by the photon energy, this transition will result in the spatial charge separation, i.e. holes in crystal and electrons in glass. Therefore, in a TiO$_{2}$ system containing both crystalline and glass phases in nanoscale, we expect that spatial charge separation\cite{N23} will help the photo-induced charge carriers to survive the strong charge recombination near surface.

\begin{figure}
\includegraphics[width=0.45\textwidth,clip]{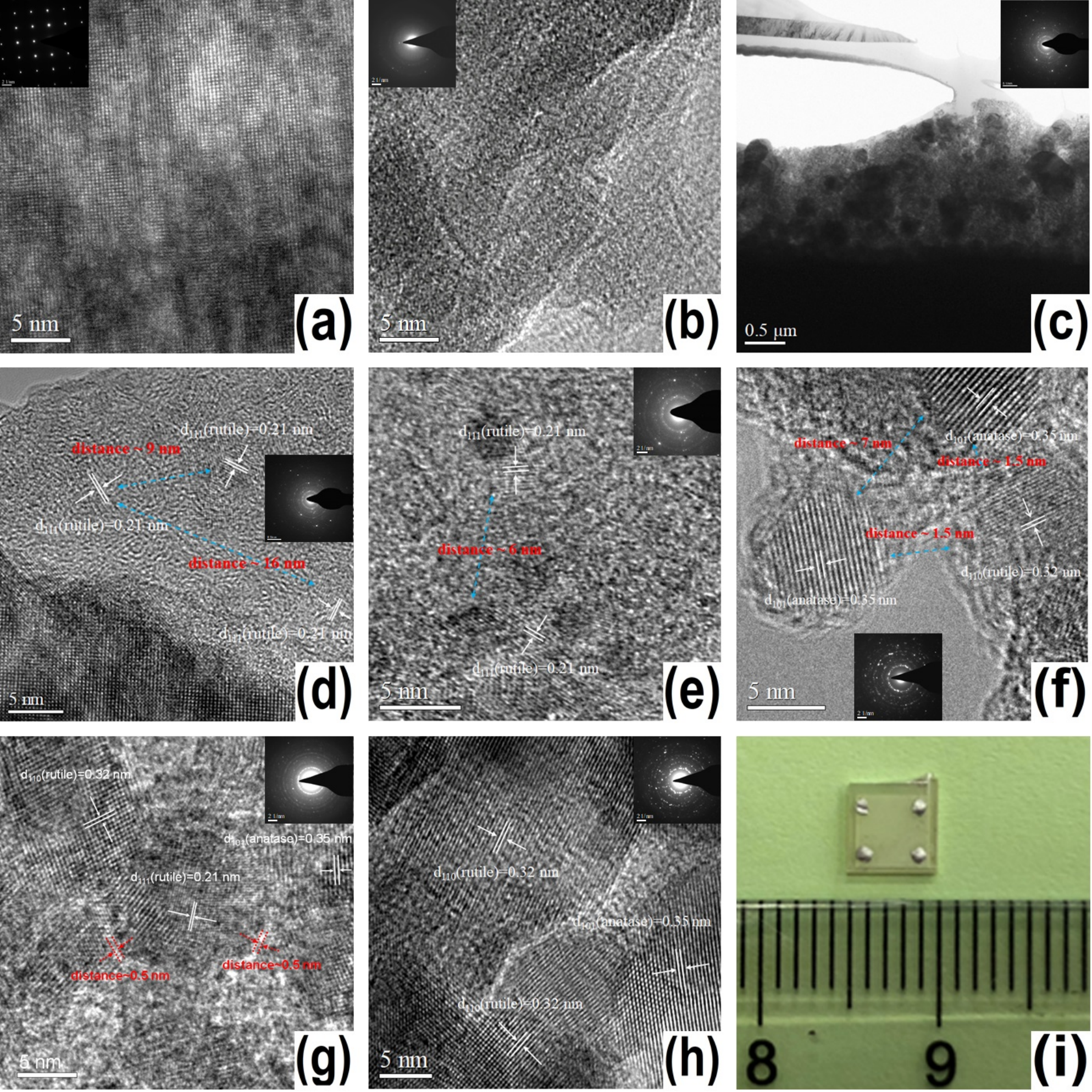}
\caption{\label{fig:figure2}The TEM images and corresponding SAED patterns of the rutile TiO$_{2}$ substrate (a), the TiO$_{2}$ glass thin film (b) produced by PLA on the rutile substrate, the 1-$mu$m thick TiO$_{2}$ glass film with nanocrystals inside (c); the TiO$_{2}$ glass films annealed slightly (d), for 1 hour (e), for 4 hours (f), for 6 hours (g), and for 8 hours (h); and the physical photo of the In electrodes for the photo-responsivity measurements (i).}
\end{figure}

To test the NanoHLET idea proposed above, we fabricated the thin layer of mixed crystalline and glassy TiO$_{2}$ in nanocale on a rutile substrate using pulsed laser ablation (PLA) and post-growth thermal annealing techniques. On a bulk rutile surface, PLA produced a glassy thin film of $\sim$1$\mu$m thick as shown by their transmission electron microscopy (TEM) images and selected area electron diffraction (SAED) pattern as shown in Figs.\ref{fig:figure2}(a), (b) and (c). By tuning the annealing time and temperature, different types of nanocrystals could be formed inside the glassy film with controlled morphology. For example, by annealing for half an hour under a constant temperature of 700$^o$C rutile nanocrystals appeared sparsely in the glass film as shown in Fig.\ref{fig:figure2}(d). They are separated from each other by about 10 nm or slightly longer. Upon further annealing, more and more nanocrystals appeared in the glassy film. Fig.\ref{fig:figure2}(e) shows the TEM image of TiO$_{2}$ glass film annealed for 1 hour, where the nanocrystals are separated by about 5 nm. Fig.\ref{fig:figure2}(f) shows the picture of nanocrystals inside a glassy film after annealing for 4 fours. By now, some of the nanocrystals are separated only by about 1-2 nm. In addition, anatase nanocrystals start to form next to the rutile ones. After annealing for 6 hours, the nanocrystals are as close to each other as 0.5 nm as shown in Fig.\ref{fig:figure2}(g). Finally, after the annealing for 8 hours, the nanocrystals of both rutile and anatase have merged together to eliminate completely the glassy phase (shown in Fig.\ref{fig:figure2}(h)).

\begin{figure}
\includegraphics[width=0.45\textwidth,clip]{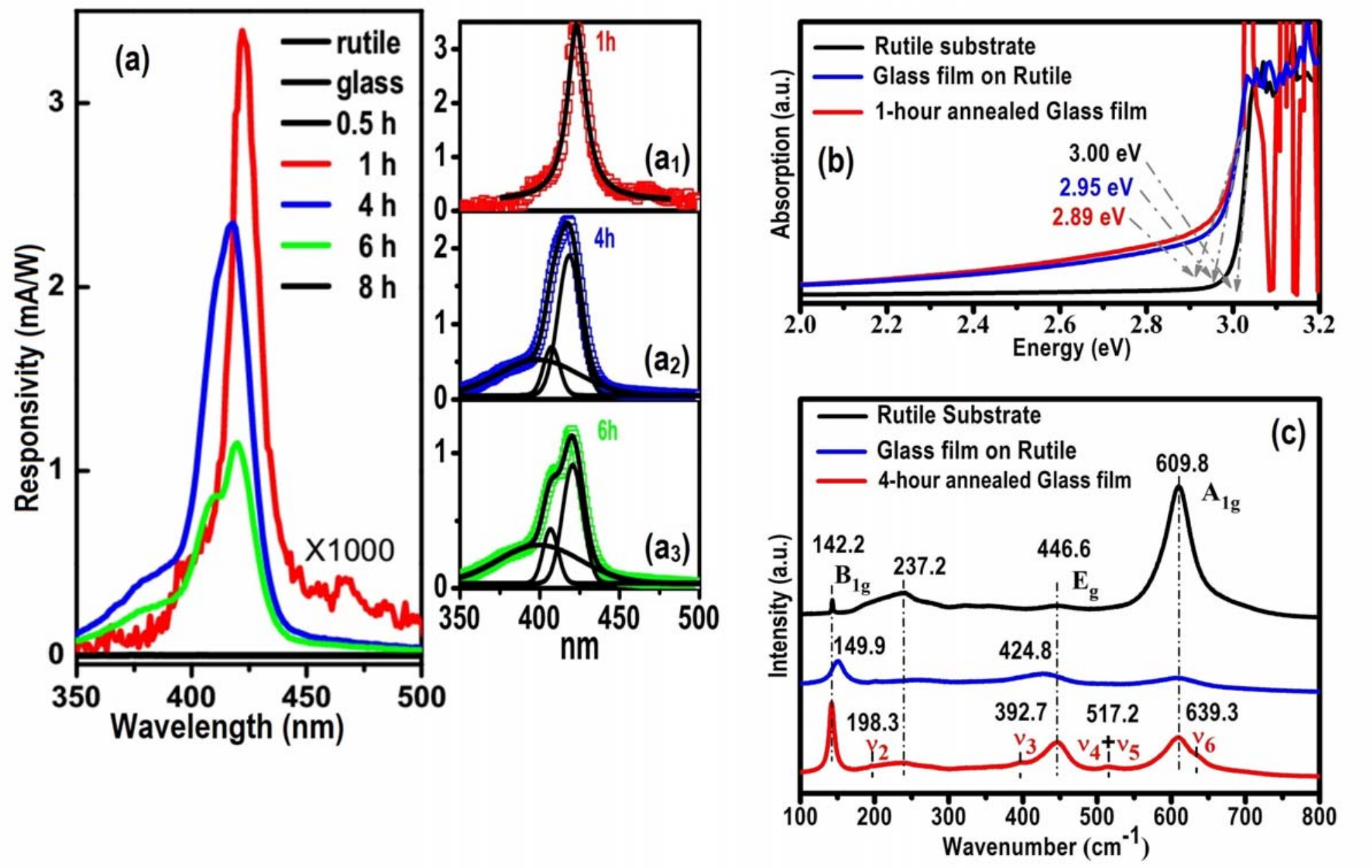}
\caption{\label{fig:figure3}The corresponding responsivity spectra of the TiO$_{2}$ films (a), and the fitted Lorentz curves of 1-hour (a1), 4-hour (a2), and 6-hour (a3) annealed films; and the absorption (b) and Raman (c) spectra of the rutile TiO$_{2}$ substrate, the glassy film and the glassy film with nanocrystals. }
\end{figure}

On top of these TiO$_{2}$ thin films, as shown in Fig.\ref{fig:figure2}(i), the indium electrodes were deposited on them, forming Ohmic contacts\cite{N24}, to collect the near-surface photocurrent directly at room temperature under the irradiation of a Xe lamp. For comparison, the bare rutile substrate and the PLA-produced glassy thin film without post-annealing were measured under the same conditions. Under a bias voltage of 40 volts applied to two electrodes separated by 3-mm, neither the rutile substrate nor the glassy film shows any measurable photoresponse for incident photons with wavelengths varying between 350 and 500 nm, as shown in Fig.\ref{fig:figure3}(a).  The optical absorption spectra of the rutile and glassy TiO2 samples were also measured as shown in Fig.\ref{fig:figure3}(b), where the crystalline rutile substrate showed an absorption edge at 3.0 eV and the glassy TiO$_{2}$ film (no annealing) with the rutile substrate showed an absorption edge at a slightly lower energy of 2.95 eV.

For the lightly annealed glassy TiO$_{2}$ film with sparsely distributed rutile nanocrystals shown in Fig.\ref{fig:figure2}(d), the same applied voltage also failed to produce any noticeable photo-response in the same spectral range as shown in Fig.\ref{fig:figure3}(a). That is not unexpected. Since in this TiO$_{2}$ film the nanocrystals are rather small and exist sparsely, the region of crystal-glass interfaces where the rutile-VBM-to-glass-CBM transition can happen takes only a small proportion in the film. Consequently, those interfaces cannot contribute enough photocurrent measurable. This situation changes drastically for the sample annealed for 1 hour. Its photo-response curve exhibits a sharp peak with a FWHM of only 13.7 nm centered at 423.0 nm (corresponding to photon energy of 2.93 eV) as shown in  Fig.\ref{fig:figure3}(a). In this case, the rutile nanocyrstals are separated by about 5 nm in the glassy matrix are much bigger as shown in Fig.\ref{fig:figure2}(e), and therefore the crystal-glass interfaces with more significant proportion can contribute enough charge carriers spatially separated that can survive the strong surface recombination and eventually form photocurrent effectively. The narrow photo-response peak shown in Fig.\ref{fig:figure3}(a) can be fitted by a Lorentzian. The fact that this peak occurs at about the same energy as the absorption edge at 2.89 eV (corresponding to 429 nm) of the same 1-hour annealed film shown in Fig.\ref{fig:figure3}(b) lends strong support to our proposal that the narrow photo-response curve can originate from the photo-excited electrons and holes from the heterolattice transitions from nanocrystals to glass.

Obviously, in the 4-hour annealed glassy film, the nanocrystals get much denser and bigger (7-8 nm in size) and the interface portion now plays a much more critical role. Accordingly, as shown in Figs.\ref{fig:figure3} (a) and (a$_2$), this film presents the peak photocurrent response of 2.35 mA/W which is larger by almost a thousand times than the response of 3.39 $\mu$A/W for the 1-hour annealed film. However, as the nanocrystals grow further as in the 6-hour annealed film, the photoresponsivity peak becomes weaker as shown in Figs.\ref{fig:figure3} (a) and (a$_3$). In this case, the nanocrystals take the overwhelming proportion in the film, and the effective photocurrent is inhibited due to the reduced heterolattice excitations at interface and the shrunk glass channels for the photo-excited electrons passing through to electrodes. Based on the same argument, we expect the photoresponsivity to further decrease when the nanocrystals start to merge together as in the 8-hour annealed film. In this sample, the responsivity spectrum does not show any sign of a peak within the same spectral range, similar to the responsivity spectrum of the crystalline rutile substrate shown in Fig.\ref{fig:figure3}(a).

It is noteworthy that both the 4-hour and 6-hour annealed films show a similar multi-peak feature in their responsivity spectra. The 4-hour annealed film possesses two sharp peaks centered at 418.5 and 407.3 nm with FWHMs of 14.1 and 10.1 nm respectively, superimposed on a broad peak centered at 399.0 nm with a FWHM of 50.0 nm, as shown in Fig.\ref{fig:figure3}(a$_2$). The 6-hour annealed film shows two sharp peaks centered at 420.5 and 406.7 nm with the FWHMs of 12.8 and 9.5 nm, respectively. These two peaks are superimposed on a broad peak centered at 400.0 nm with a FWHM of 51.9nm as shown in Fig.\ref{fig:figure3}(a$_3$). The narrow higher-energy peaks of these two films presumably originate from the photo-excitations from glass to anatase nanocrystals while the narrow lower-energy peaks in both films originate from the rutile ones. The separation of about 0.1 eV between these two sharp peaks is consistent with the CBM of anatase being higher than that of rutile by about 0.1 eV. It is also consistent with electrochemical impedance measurements which indicate that the flatband potential of the anatase (101) surface is 0.2 eV below that of the rutile (001) surface\cite{N25}. The appearance of the broad peak around 400.0 nm may suggest that a mixed `alloy' of rutile and anatase nanocrystals starts to form in the 4-hour and 6-hour annealed films. The additional disorder presented in these mixed phase nanocrystals should be responsible for the larger FWHM of their responsivity spectra.

Finally, we have also determined the structures of the nanocrystals in our films by measuring their Raman spectra which are shown in Fig.\ref{fig:figure3}(c). The substrate Raman spectrum shows peaks at 609.8, 446.6 and 142.2 cm$^{-1}$. These have been identified as the A$_{1g}$, E$_{g}$ and B$_{1g}$ modes, respectively of rutile in the literatures \cite{N26,N27,N28}. A broad peak at 237.2 cm$^{-1}$ has been identified as a two-phonon mode. After PLA, the sharp Rutile Raman peaks were replaced by three broad peaks around 609.8, 424.8 and 149.9 cm$^{-1}$ as a result of vitrification. After annealing for 4 hours, recrystallization caused the three Rutile peaks to become sharper again. At the same time new modes appeared at 198.3, 393.7, 517.2, and 639.3 cm$^{-1}$. These new peaks have been attributed to Raman modes of anatase designated as $\nu$$_2$, $\nu$$_3$, $\nu$$_4$+$\nu$$_5$, and $\nu$$_6$, respectively in the literature\cite{N26}.

In single phases, normal TiO$_{2}$ thin films have been found to be broad-band photoconductors which show cutoff wavelengths of about 380, 380, 385 and 400 nm and exhibit maximum responsivities of about 5.0, 3.0, 2.5, and 1.5 A/W, as in the amorphous, anatase, mixed anatase/rutile, and rutile phases, respectively\cite{N29}. These results were measured using the interdigitated electrodes with spacing of 10 $\mu$m and bias voltage of 5 V. Considering the fact that the photoconductive gain is normally proportional to the bias voltage and inversely proportional to the square of the electrode spacing\cite{N30,N31}, the 4-hour annealed thin film, in its point-electrodes configuration with spacing of 3 mm and bias voltage of 40 V, actually exhibits the photoconductive performance as good as these broad-band TiO$_{2}$ films can do.

In conclusion, if engineered properly, the nanoscale crystal-glass interfaces in TiO$_{2}$ accommodate effectively the heterolattice electronic transitions that leads to the spatially separated charge carriers under photon irradiation. With the simple photoconductive measurements between point electrodes, these particular carriers have been demonstrated able to survive the strong recombination near surface and result in the ultra-sharp photoresponse with FWHM as small as 13.7 nm. Distinct to the previous ones, the NanoHLET is a near-surface mechanism for narrow-band photodetection that does not require any near-surface region or extra surface features for realizing frequency selection. This mechanism makes it possible to fabricate the much smaller narrow-band photodetection pixels in submicron scale and consequently to achieve the high-density detector arrays using current semiconductor nanotechnology.  In principle, the NanoHLET mechanism should also be applicable to other semiconductors, especially metal oxides.

\begin{acknowledgments}
This work is supported by the National Natural Science Foundation of China (11727902, 12074045, 11674038, 11774341 and 12074372), the National Science Fund for Distinguished Young Scholars of China (No. 61525404), the Open Project of the State Key Laboratory of Luminescence and Applications, the Foundation of State Key Laboratory of High Power Semiconductor Lasers, and the Developing Project of Science and Technology of Jilin Province (20160519007JH, 20160520117JH).
\end{acknowledgments}


\end{document}